\documentstyle[aps,epsf]{revtex}
\newcommand{\be}{\begin{equation}}
\newcommand{\ee}{\end{equation}}
\newcommand{\bea}{\begin{eqnarray}}
\newcommand{\eea}{\end{eqnarray}}
\newcommand{\nn}{\nonumber \\}
\newcommand{\eqn}[1]{(\ref{#1})}
\font\mybb=msbm10 at 10pt
\def\bb#1{\hbox{\mybb#1}}
\def\bZ {\bb{Z}}

\def\bE {\bb{E}}

\def\bC {\bb{C}}

\def\bfsig{\mbox{\boldmath $\sigma$}}

\def\nax {{\bf \nabla} X}
\def\nay {{\bf \nabla} Y}

\def\a {\alpha}
\def\b {\beta}
\def\c {\gamma}

\def\pa{\partial}

\begin{document}

\twocolumn[\hsize\textwidth\columnwidth\hsize\csname
@twocolumnfalse\endcsname

\rightline{DAMTP-1999-38}
\rightline{ECM-UB-99/07}
\rightline{QMW-PH-99-05}
\rightline{hep-th/9903156}

\title{Finite energy Dirac-Born-Infeld monopoles and string junctions} 
\author{J.P. Gauntlett$^a$, ~C. Koehl$^a$, ~D. Mateos$^b$, ~P.K. Townsend$^{c}$
and M. Zamaklar$^{c}$\\}
\address{
$^a$
Department of Physics, Queen Mary and Westfield College,\\
University of London, Mile End Road,\\
London E1 4NS, U.K. \\
$^b$
Departament ECM, Facultat de F{\'\i}sica, \\
Universitat de Barcelona and Institut de F{\'\i}sica d'Altes Energies,\\
Diagonal 647, E-08028 Barcelona, Spain\\
$^c$ 
DAMTP, University of Cambridge, Silver Street,\\
Cambridge CB3 9EW,  UK\\
}
\maketitle
\begin{abstract}
It is shown that the worldvolume field theory of a single D3-brane in a
supergravity D3-brane background admits finite energy, and
non-singular, abelian monopoles and dyons 
preserving 1/2 or 1/4 of the ${\cal N}=4$
supersymmetry and saturating a Bogomol'nyi-type bound. The 1/4 supersymmetric
solitons provide a worldvolume realisation of string-junction dyons. We also
discuss the dual M-theory realisation of the 1/2 supersymmetric dyons as
finite tension self-dual strings on the M5-brane, and of the 1/4
supersymmetric dyons as their intersections.  
\end{abstract}
\vskip2pc]

\section{Introduction}

The ${\cal N}=4$ supersymmetric Yang-Mills (SYM) 
theory with gauge group $SU(k)$
spontaneously broken to $U(1)^{k-1}$ has a spectrum 
of 1/2 supersymmetric magnetic
monopoles and dyons which, together with the `elementary' particles of
the perturbative spectrum, fill out orbits of an $SL(2;\bZ)$ electromagnetic
duality group. Each such particle has an interpretation in IIB superstring
theory as an $(m,n)$ string stretched between a pair of parallel D3-branes,
chosen from among $k$ parallel D3-branes. For $k=2$ there are 
no other particles
in the spectrum but for $k\ge3$ there are additional, 1/4 supersymmetric, 
dyons that
are entirely non-perturbative in the sense that they belong to $SL(2;\bZ)$
orbits that  contain no `elementary' particles. Although these can be found as
classical solutions of the SYM field equations \cite{hhs,ko,lee} 
they were first found
as IIB superstring configurations in which three strings of different $(m,n)$
charges, attached to three D3-branes, meet at a string junction \cite{berg}.
These are points at which two IIB strings of charges $(m,n)$ and $(m',n')$ meet
to form a string of charge $(m+m',n+n')$ \cite{schwarz}. The minimum energy 
state to which the configuration relaxes is one in which three strings leaving
the three D3-branes meet at a planar string junction \cite{mukhi,GGT}.

Actually, the effective action of the D3-branes 
is not a SYM theory but rather a
supersymmetric non-abelian Dirac-Born-Infeld (DBI) theory. The precise nature 
of this theory is not known (see \cite{brecher} for a recent discussion) but it
has an expansion in powers of $\alpha' {\cal F}$ that simplifies in certain
limits; $\alpha'$ is the inverse IIB string tension and ${\cal F}$ is
the (background covariant) Born-Infeld field strength. If $L$ is the minimal
separation between the D3-branes then (as we shall later see explicitly)
$\alpha' {\cal F}\sim L^2/\alpha'$, so the expansion parameter is actually
$L^2/\alpha'$. For $L<<\sqrt{\alpha'}$ we need keep only the quadratic terms in
${\cal F}$ and the action reduces to the ${\cal N}=4$ SYM theory (for a vacuum IIB
background). For $L>>\sqrt{\alpha'}$ we cannot truncate 
the expansion but we may
neglect the non-abelian interactions; the action then reduces to a sum of
abelian DBI actions governing the dynamics of independent parallel D3-branes.
The D3-brane action depends on the supergravity background. For example,
${\cal F}= F-B$ where $F$ is the usual 2-form $U(1)$ field strength, satisfying
$dF=0$, and $B$ is the pullback of the background NS-NS 2-form potential. The
D3-brane couples to the background R-R gauge fields through a Wess-Zumino
Lagrangian $L_{WZ}$. Let $\xi^i$ be the worldvolume 
coordinates, $(i=0,1,2,3)$. Omitting
fermions and setting $\alpha'=1$, the Lagrangian is then
\be
L = - e^{-\phi} \sqrt{-\det(g_{ij} + {\cal F}_{ij})} + L_{WZ}
\ee
where $g_{ij}$ is the induced worldvolume metric and $\phi$ the
background dilaton field. For the backgrounds we 
consider, $\phi$ is a constant,
$B$ vanishes, and $L_{WZ}$ is just the minimal coupling of the D3-brane to the
4-form gauge potential of IIB supergravity. 

For widely separated branes it makes sense to ask what a dyon looks
like locally on one of the D3-branes, {\it i.e.} as a 
solution of the {\it abelian} ${\cal N}=4$ supersymmetric DBI theory with the
above Lagrangian. Because abelian monopoles and dyons 
have infinite energy, this
question would not make sense in the context of a $U(1)$ SYM theory, but the
infinite energy has a natural interpretation in the DBI context as the energy
associated with an infinite string of fixed tension. In fact, the abelian DBI
theory does have infinite energy 1/2 supersymmetric solutions that appear as
`spikes' on the worldvolume with uniform energy per unit length \cite{CM,Gib}.
The `spike' solutions of the abelian DBI theory 
were called BIons in \cite{GGT},
following a slightly different use of this term in
\cite{Gib}.  In the case of the D3-brane there are dyonic $(m,n)$-BIons
corresponding to infinite $(m,n)$ strings that end on the D3-brane. 
Although the infinite energy of a BIon has a clear physical interpretation, it 
is nevertheless a cause for concern because, 
for example, solutions with infinite
energy  make no contribution to the semi-classical evaluation of the
path-integral.  One should really think of these solutions as limiting cases of
the more physical situation in which the string eventually ends on another
D3-brane, but it might then appear that  we are forced to return to the
non-abelian DBI theory. One purpose of this paper is to show that this problem
can be circumvented by replacing the second D3-brane by a supergravity D3-brane
background. We remark that the infinite energy problem is also 
cirumvented by certain {\it non-supersymmetric} solutions of the DBI
action \cite{Gib}, but these do not correspond to BPS states of ${\cal
N}=4$ SYM theory.   

The super D3-brane action can be consistently formulated in any background that
solves the equations of IIB supergravity. One 
such solution is the `supergravity
D3-brane'. For this solution the dilaton is  constant and we shall set it to
zero; this corresponds to unit string coupling constant. The remaining
non-vanishing fields are the metric and the 4-form potential $C$ with self-dual
5-form field strength $R=dC$. These are given by
\bea\label{back}
ds^2 &=& H^{-1/2}ds^2(\bE^{(1,3)}) + H^{1/2} ds^2(\bE^6)\nn
R &=& vol(\bE^{(1,3)})\wedge dH^{-1} + \star_6 dH
\eea
where $\star_6$ is the Hodge dual on $\bE^6$ and $H$ is harmonic on this space.
Point singularities of $H$ are coordinate singularities of the spacetime metric
at connected components of a degenerate event horizon. The proper distance to
the horizon on spacelike hypersurfaces of constant $\bE^{(1,3)}$ coordinates is
infinite, so that there are `internal' asymptotic infinities. If we wish the
D3-brane horizon to have a single connected component then we must choose a
`single-centre' metric with
\be\label{harm}
H= 1 + {Q\over |\vec{X}-\vec{X}_0|^4}~,
\ee
where $\vec{X}$ are cartesian coordinates on $\bE^6$ and $\vec{X}_0$ is a
constant $\bE^6$ 6-vector. Let us now put a test D3-brane in 
this background, at
$\vec{X}=0$. The DBI equations can now have solutions representing infinite
$(m,n)$  strings that go into the internal asymptotic region of the 
background geometry. We shall show that there exist static BIon solutions of
this type, and that they have a {\it finite} energy, saturating a
Bogomol'nyi-type bound. In fact, if $|\vec{X}_0|=L$ then the energy of the 
static BIon is precisely $L$ times the tension of 
an $(m,n)$ string. Effectively,
we have replaced the `second' D3-brane of the $SU(2)$ theory by a D3-brane
background, thereby finding finite energy, and 
non-singular, supersymmetric monopoles and dyons
in the {\it abelian} DBI theory. Actually, it 
would be more accurate to consider
this `brane in brane background' configuration as representing the large $k$
limit of an $SU(k)$ theory broken 
to $SU(k-1)\times U(1)$ with the $SU(k-1)$ theory
replaced by the supergravity background. 

The same logic that leads us to expect 1/2 supersymmetric BIons on the D3-brane
also leads us to expect that it should be possible to find the 1/4
supersymmetric string junctions this way. Consider first the case in which one
of the three strings in a string junction configuration has shrunk to zero
length. In this case we are left with a configuration of two `overlapping'
strings of different $(m,n)$ charges each stretched between a different pair of
D3-branes. In the special case of an orthogonal overlap of an F-string, 
charge (1,0), with a D-string, charge (0,1), this configuration can be
represented by the array 
$$
\begin{array}{lccccccccc}
D3: & 1 & 2 & 3 & - & - & - & - & - & - \nn
D1: & - & - & - & 4 & - & - & - & - & - \nn
F1: & - & - & - & - & 5 & - & - & - & - 
\end{array}
$$ 
The corresponding 1/4 supersymmetric dyon solution on the D3-brane was
recently found \cite{GP}; it depends on two independent worldspace functions
that are harmonic in the Euclidean metric. Here we shall find the general
`two-harmonic-function' solution and explain its interpretation as a
string junction. In a flat background these solutions again have infinite 
energy. The strategy explained above to find finite energy solutions can be
used here too, but in this case we must use a background harmonic function $H$
with {\it two} isolated singularities, {\it i.e.} we replace (\ref{harm}) 
by
\be
H = 1 + {Q_1\over |\vec{X}-\vec{X}_1|^4} + {Q_2\over |\vec{X}-\vec{X}_2|^4}
\ee
where $\vec{X}_1$ and $\vec{X}_2$ are two 6-vectors 
giving the positions of the 
background supergravity D3-branes. A string leaving the test D3-brane can now
split, at a string junction, into two strings, each of which continues
indefinitely into one of the two `internal' asymptotic regions of the
background geometry. As we shall see, such configurations correspond to
{\it finite} energy abelian DBI solitons saturating precisely the
Bogomol'nyi-type bound expected of a 1/4 supersymmetric dyon. 

The 1/2 supersymmetric dyons on a D3-brane of IIB superstring theory
have an M-theory counterpart as self-dual string solitons on an
M5-brane \cite{HLW,GGT}. In flat D=11 spacetime these strings have
infinite tension, as expected from their spacetime interpretation as
semi-infinite M2-branes with a boundary on an M5-brane. However
the M5-brane action can be consistently formulated in any background
that solves the equations of
D=11 supergravity. One such solution is the supergravity M5-brane. The
11-metric and 4-form field strength of this solution are
\bea
ds^2_{11} &=& U^{-1/3}ds^2(\bE^{(5,1)}) + 
U^{2/3}d\vec{X}\cdot d\vec{X}\nn
F_{(4)} &=& \star_5 dU
\label{m5sol}
\eea
where $\vec{X}$ are cartesian coordinates on $\bE^5$ and $U$ is a
harmonic function on this space. Singularities of $U$ are just
horizons of the 11-metric which are at an infinite proper distance on
the spacelike hypersurfaces of constant $\bE^{(5,1)}$ coordinates.
In other words, there are again `internal' asymptotic regions into
which we can take an M2-brane emanating from a test M5-brane in this
background. In this way we find self-dual string solitons on the M5-brane
worldvolume with {\it finite} tension.
The 1/4 supersymmetric dyons on the IIB D3-brane also have an
M-theory analogue, this time as intersecting self-dual string
solitons on the M5-brane. For an M5-brane in flat spacetime, these
were found in \cite{GLW} from the requirement of 1/4 supersymmetry; 
here we show that they saturate a Bogomol'nyi-type bound,
although the total energy is, of course, infinite. By considering the
M5-brane in a two-centre M5-brane background we are able to find
solutions that represent intersecting finite-tension self-dual
strings.

We begin with some details of the D3-brane Hamiltonian in a
supergravity D3-brane background that we need for our subsequent discussion of
finite energy BIons and string junctions. We then discuss the dual
M-theory realisation of these solitons.

\section{D3-brane Hamiltonian in a D3-brane background}
\label{d3ham}

The Hamiltonian form of the super D3-brane Lagrangian density in a general
superspace background was given in \cite{BT}. Setting fermions to zero and
specializing to a background of the form assumed above we have
\be
{\cal L} = P_m \dot X^m + E^a \dot V_a + V_t {\cal G}
- s^a {\cal P}_a - {1\over2}v {\cal H}
\ee
where $X^m$ are the spacetime coordinates and $P_m$ the
10-momentum $(m=0,1,\dots,9)$, and $V_a$ is the BI 3-vector potential 
$(a=1,2,3)$ and $E^a$ its conjugate electric field 3-vector. The constraint
functions associated with the constraints imposed by the Lagrange multiplers
$V_t$, $s^a$ and $v$ are
\bea
{\cal H} &=& G^{mn}\left(P - {\cal C}\right)_m\left(P -
{\cal C}\right)_n + E^aE^bg_{ab} + \det\left(g+ F\right) \nn
{\cal G} &=& \pa_a E^a \nn
{\cal P}_a &=& \left(P - {\cal C}\right)_m\pa_a X^m + E^bF_{ba}
\eea
where $G$ is the background metric,
$g$ is the induced worldspace metric, $F$ the magnetic field 2-form and
${\cal C}_m$ is the coefficient of $\dot X^m$ in the WZ term, i.e $L_{WZ}=\dot
X^m {\cal C}_m$.

In the static gauge 
\be
X^m = (\xi^i, \vec{X})\, , 
\ee
where $\xi^i=(t,\bfsig^a)$ are the worldvolume coordinates, we have
\be\label{cee}
{\cal C}_m =({\cal C}, 0, \vec{C})\, ,
\ee
where, for a background of the form assumed here, 
\be
{\cal C}= H^{-1}\, .
\ee
In addition, the constraint ${\cal P}_a=0$ implies, in static gauge, that
\be
P_m = \left(-{\cal E}, - 
(\vec{P}-\vec{C})\cdot \pa_a \vec{X} -E^bF_{ab},\, \vec{P}\right)\, 
\ee
where ${\cal E}$ is the energy density. Since $\vec{P}-\vec{C}$ vanishes for
static configurations we have
\be
(P-{\cal C})_m = \left({-\cal E}- H^{-1},-E^bF_{ab},\, \vec{0}\right)\, .
\ee
If this is now used in the constraint ${\cal H}=0$ we can solve for ${\cal
E}$ to get
\bea
({\cal E} + H^{-1})^2 &=& E^cE^dF_{ac}F_{bd}\delta^{ab} \nn
&+& H^{-{1\over2}}\left[E^aE^b g_{ab} + \det(g+F)\right]
\eea
where
\be
g_{ab} = H^{-{1\over2}}\delta_{ab} + 
H^{1\over2}\pa_a\vec{X}\cdot \pa_b\vec{X}~.
\ee
When $H=1$ this reduces to the result given in \cite{GGT} except for a shift of
the vacuum energy. To obtain precisely the result of \cite{GGT} when $H=1$ one
would have to take ${\cal C}=H^{-1}-1$; the difference is just a gauge
transformation and hence without physical significance. The choice ${\cal
C}=H^{-1}$ is convenient because it ensures that the WZ term cancels the vacuum
energy of the DBI term.  

\section{Finite energy BIons}
\label{bions}

We shall begin by choosing the harmonic function $H$ to have the form
(\ref{harm}) with $\vec{X}_0=(L,0,\dots,0)$. On the $X$-axis we
then have
\be
H= 1+{Q\over (X-L)^4} \, .
\ee
We shall be interested in BIons that can be interpreted as strings stretched
between the test D3-brane and the source D3-brane. It is obvious that the
minimum energy configuration must then be one for which the only non-zero
worldvolume field is $X(\bfsig)$ so we now set the others to zero.
In this case 
\be
g_{ab} = H^{-{1\over2}}\delta_{ab} + H^{1\over2}\pa_a X\pa_b X
\ee 
and 
\bea
\det(g+F) &=& H^{-{3\over2}}\left[ 1+ H|{\bf \nabla} X|^2 + H|{\bf B}|^2\right] \nn
&& +\, H^{1\over2}({\bf \nabla} X\cdot {\bf B})^2
\eea
where ${\bf B}$ is the magnetic field 3-vector defined by 
\be
F_{ab}=\varepsilon_{abc}B_c\, .
\ee
This leads to the formula
\bea
({\cal E} + H^{-1})^2 &=& H^{-2} + H^{-1}\left[|{\bf \nabla} X|^2 + |{\bf E}|^2 +
|{\bf B}|^2\right] \nn
&& + ({\bf \nabla} X\cdot {\bf E})^2 
+({\bf \nabla} X\cdot {\bf B})^2 + |{\bf E}\times {\bf B}|^2
\eea
which we can rewrite, for arbitrary angle $\vartheta$, as
\bea
({\cal E} + H^{-1})^2 &=& \left[H^{-1} + 
\cos\vartheta ({\bf E}\cdot {\bf \nabla} X) + \sin\vartheta ({\bf B}\cdot {\bf
{\bf \nabla}}X) \right]^2 \nn
&+& H^{-1}|{\bf E} -\cos\vartheta {\bf \nabla} X|^2 
+ H^{-1}|{\bf B} - \sin\vartheta {\bf \nabla} X|^2 \nn
&+& |\sin\vartheta ({\bf E}\cdot {\bf \nabla} X) - \cos\vartheta 
({\bf B}\cdot {\bf \nabla} X)|^2 \nn
&+& |{\bf E}\times {\bf B}|^2\, .
\eea
{}From this expression we deduce the bound
\be
({\cal E} + H^{-1})^2 \ge \left[ H^{-1} + 
\cos\vartheta ({\bf E}\cdot {\bf \nabla} X) + \sin\vartheta ({\bf B}\cdot {\bf
{\bf \nabla}} X\right]^2~.
\ee
and hence that
\be
{\cal E} \ge \cos\vartheta ({\bf E}\cdot {\bf \nabla} X) +
\sin\vartheta ({\bf B}\cdot {\bf {\bf \nabla}} X)
\ee
for any $\vartheta$. Integrating over the worldspace and then
maximising the right hand side with respect to $\vartheta$ we arrive
at the bound
\be\label{bound}
M \equiv \int d^3 \sigma \, {\cal E} \ge \sqrt{Q_e^2 + Q_m^2}
\ee
where 
\bea\label{surf}
Q_e &=& \int d^3\sigma\, {\bf E}\cdot {\bf \nabla} X = 
\oint d{\bf S}\cdot {\bf E} X~, \nn
Q_m &=& \int d^3\sigma\, {\bf B}\cdot {\bf \nabla} X =
\oint d{\bf S}\cdot {\bf B} X~.
\eea
Here we have used the fact that ${\bf B}$ is divergence-free by
definition and that ${\bf E}$ is divergence-free as a result of the
Gauss law constraint.
The final surface integrals must be taken over all components of the
boundary surface of worldspace.

The bound (\ref{bound}) is saturated by solutions of
\be\label{rein}
{\bf E} = \cos\vartheta {\bf \nabla} X\,,\qquad
{\bf B} = \sin\vartheta {\bf \nabla} X~.
\ee
where $\tan\vartheta =Q_m/Q_e$.
These are just the flat space abelian Bogomol'nyi equations. Since both ${\bf
E}$ and ${\bf B}$ are divergence-free we deduce, as in the flat-space 
case, that $X$ is harmonic on worldspace (in the Euclidean metric). 
For a single BIon of charge $q$ with $X$ vanishing at spatial infinity on
the brane we have
\be\label{exx}
X= q/4\pi r \qquad (r > r_0 \equiv q/4\pi L)
\ee
where $r=|\bfsig|$ is the distance from the origin in $\bE^3$. 

The lower bound on $r$ comes about because the worldspace metric is
\be
ds^2_3 = \left[ H^{-{1\over2}} + \left({Lr_0\over r^2}\right)^2
H^{1\over2}\right] dr^2 + H^{-{1\over2}}d\Omega_2^2\, ,
\ee
where now
\be
H = 1 + {Qr^4\over L^4(r-r_0)^4}\, .
\ee
As $r$ approaches $r_0$ from above we have
\be\label{nearsing}
ds^2_3 \sim Q^{1\over2}\left({du\over u}\right)^2 + Q^{-{1\over2}}L^2 u^2
d\Omega_2^2 
\ee
where $u=r-r_0$. The proper distance to $r=r_0$ is therefore infinite. In fact,
the sphere at $r=r_0$ is mapped to a single point $\vec{X}=\vec{X}_0$ in the
transverse space. The worldspace of a single BIon along the 
X-axis therefore has
two boundaries: one at $r=\infty$, where $X=0$, and another 
at $r=r_0$, where $X=L$. The surface integrals of (\ref{surf}) vanish 
at the $r=\infty$ boundary, 
and since $X=L$ on the other boundary we have 
\be
Q_e = L\,q_e\, , \qquad Q_m = L \, q_m\, ,
\ee
where
\bea
q_e &=& \oint d{\bf S}\cdot {\bf E} = \cos\vartheta \, q~,\nn
q_m &=& \oint d{\bf S}\cdot {\bf B} = \sin\vartheta \, q
\eea
are the electric and magnetic charges coupling to the BI field on the brane.
The BIon mass is therefore
\be
M= |q|L/(\alpha')^2
\ee
where we have now reinstated $\alpha'$. This
is the mass of a string of tension $|q|/(\alpha')^2$ and length $L$.
Taking into acount the quantization condition on the IIB string
charges, and the fact that we have set the string coupling constant to
unity, the tension of an $(m,n)$ string is 
$\sqrt{m^2 +n^2}/\alpha'$, so
\be
q= \alpha'\sqrt{m^2 + n^2}\, .
\ee
Reinstating $\alpha'$ in (\ref{rein}) we see that the maximum value of
$\alpha' E$ or $\alpha'B$ is $L^2/q \sim L^2/\alpha'$, so the
expansion parameter of the DBI action is $L^2/\alpha'$, as claimed
earlier. Actually, if $m$ or $n$ is very large then the effective 
expansion parameter is really smaller than $L^2/\alpha'$ by a factor
of $1/\sqrt{m^2+n^2}$. For given $L$, the neglect of the DBI corrections to
the SYM theory can therefore be justified by considering a sufficiently 
large charge, so the SYM theory is adequate for a description of
macroscopic objects. The DBI corrections are important only
for the description of microscopic objects. 

In terms of the electrostatic analogy, the above construction can be viewed
as a regularization of the infinite self-energy of a point particle in which a
point charge is replaced by a perfectly conducting charged spherical shell. 
This is clearly unsatisfactory as a solution to the electrostatic self-energy
problem of electrodynamics because any surface of spherical topology carrying 
the same total charge would serve the same purpose. Here too we could replace
the spherical shell by a shell of any other shape, but in our case this has no
effect on the physics. To see this we first note that the surface of the shell
is, by hypothesis, an equipotential with potential $X=L$. It follows that every
point on it is mapped to the point $\vec{X}=\vec{X}_0$ in 
transverse space. This
point is at infinite proper distance in the transverse space metric and hence
in the induced worldspace metric. The equipotential surface $X=L$ in $\bE^3$ is
therefore a point at infinity in the induced worldspace
metric. Neighbouring equipotential surfaces of constant $X<L$ can be
used to define coordinates in the neighbourhood of $X=L$ for which $X$
is again given by (\ref{exx}). It then
follows that the induced worldvolume metric in this neigbourhood is
(\ref{nearsing}). But the minimum energy metric is determined by its behaviour
near points at infinity. The initial shape of the shell is therefore 
irrelevant to the final solution. 
 
The spherical symmetry of the one BIon solution is therefore in no way
essential to the construction of finite energy BIons, and multi-BIon solutions
can be constructed analogously: we remove $n$ closed surfaces of
spherical topology from $\bE^3$  and choose their 
potential $X$ to correspond to
the centre of the background D3-brane spacetime metric. 
The potential is set to zero at
infinity. There is now a unique solution of Laplace's equation for $X$; this is
the sought multi-BIon solution. Using this solution we may compute the
worldspace metric to which it corresponds. This metric will have $n$ points at
infinity, near each of which it will take the  form (\ref{nearsing}). Different
shapes of the initial surfaces just correspond to different choices of
coordinates for the (unphysical) Euclidean 3-space.

\section{DBI string junctions}

In this section we will first derive the general bound
saturated by 1/4 supersymmetric dyons in a general two-centre 
D3-brane background; then we will  
move on to discuss their interpretation as string
junctions in the simpler case of a flat background,
to return finally to their interpretation in the general case.

\subsection{The BPS bound}
So far we have considered a test D3-brane in a single-centre
parallel D3-brane background. We now want to consider a two-centre background,
e.g
\be
H= 1 + {1\over |\vec{X} - \vec{X}_1|} + {1\over |\vec{X} - \vec{X}_2|}
\label{Hwith2centres}
\ee
where $\vec{X}_1$ and $\vec{X}_2$ are two non-zero vectors. We may choose them
to lie in the plane for which $\vec{X}=(X,Y,0,\dots,0)$. A static
D3-brane configuration in such a background will generally have non-constant
$X(\bfsig)$ and $Y(\bfsig)$. The same reasoning as before now
leads to the following expression for the energy density:
\bea
\left({\cal E} + H^{-1}\right)^2 &=& H^{-2} + H^{-1}\left[
|{\bf \nabla} X|^2 + |{\bf \nabla} Y|^2 + |{\bf E}|^2 + |{\bf B}|^2\right]\nn
&+&  ({\bf \nabla} X\cdot {\bf E})^2 + ({\bf \nabla} Y\cdot {\bf E})^2\nn
&+& ({\bf \nabla} X\cdot {\bf B})^2 +  ({\bf \nabla} Y\cdot {\bf B})^2 \nn
&+& |{\bf E}\times {\bf B}|^2 + |{\bf \nabla} X \times {\bf \nabla} Y|^2\,.
\eea
We can rewrite the right hand side, for arbitrary angle $\alpha$ as
\bea
&[ H^{-1} + \cos\alpha\, {\bf E}\cdot
{\bf \nabla} X - \sin\alpha\, {\bf E}\cdot{\bf \nabla} Y  \nn
&\qquad\qquad\qquad  + \sin\alpha\, {\bf B}\cdot{\bf \nabla} X
+ \cos\alpha\, {\bf B}\cdot{\bf \nabla} Y]^2\nn
&+ H^{-1}\left[ {\bf E} -\cos\alpha\, {\bf \nabla} X + \sin\alpha\, {\bf \nabla}
Y\right]^2\nn &+ H^{-1}\left[ {\bf B} -\sin\alpha\, {\bf \nabla} X - \cos\alpha\,
{\bf \nabla} Y\right]^2\nn &+ [\sin\alpha\, {\bf E}\cdot{\bf \nabla} X + \cos\alpha\, {\bf
E}\cdot{\bf \nabla} Y\nn &\qquad\qquad\qquad - \cos\alpha\, {\bf B}\cdot{\bf \nabla} X + 
\sin\alpha\, {\bf B}\cdot{\bf \nabla} Y]^2\nn
&+ |{\bf E}\times {\bf B} - {\bf \nabla} X \times {\bf \nabla} Y|^2\,.
\eea
We thereby deduce that
\bea\label{twoscalarbound}
&{\cal E} \ge \cos\alpha\, {\bf E}\cdot
{\bf \nabla} X - \sin\alpha\, {\bf E}\cdot{\bf \nabla} Y \nn
& \qquad\qquad + \sin\alpha\, {\bf B}\cdot{\bf \nabla} X
+ \cos\alpha\, {\bf B}\cdot{\bf \nabla} Y\, ,
\eea
for any $\alpha$.
By integrating over the worldspace and maximising the
right hand side with respect to $\alpha$, 
we deduce that the total mass $M$
satisfies the bound
\be\label{bbound}
M \ge \sqrt{(Q_m^X - Q_e^Y)^2 + (Q_m^Y + Q_e^X)^2} 
\ee
where $(Q_e^X,Q_e^Y)$ and $(Q_m^X,Q_m^Y)$ are the non-vanishing components of 
the electric and magnetic charge 6-vectors
\be
\vec{Q}_e = \oint d{\bf S}\cdot {\bf E} \vec{X}\,, 
\qquad 
\vec{Q}_m = \oint d{\bf S}\cdot {\bf B} \vec{X}\, .
\ee
The bound is saturated when
\bea\label{equality}
{\bf E} &=& \cos\alpha {\bf \nabla} X -\sin\alpha {\bf \nabla} Y\,, \nn
{\bf B} &=& \sin\alpha {\bf \nabla} X +\cos\alpha {\bf \nabla} Y\,\, ,
\eea
where
\be
\tan\alpha = {Q_m^X - Q_e^Y\over Q_e^X + Q_m^Y}\, .
\ee

The mass $M$ of configurations saturating the bound may be 
rewritten in the form
\be\label{bpse}
M\ge \sqrt{|\vec{Q}_e|^2 + |\vec{Q}_m|^2 + 2 \, |\vec{Q}_e|
\,|\vec{Q}_m|\, \sin\xi }
\ee
where $\xi$ is the angle between the two 6-vectors $\vec{Q}_e$ and
$\vec{Q}_m$.  
This is precisely the mass formula for $1/4$ supersymmetric
dyons in ${\cal{N}}=4$  $D=4$ SYM theories \cite{hollowood,berg}. It is
invariant under $SO(6)$ rotations of the 6-vectors $\vec{Q}_e$ 
and $\vec{Q}_m$, and 
under an $SO(2)$ rotation of these two 6-vectors into each other, as
expected from the $U(4)$ automorphism group of the ${\cal N}=4$ 
$D=4$ supersymmetry
algebra. When $\sin\xi=0$ we recover the  formula for 1/2
supersymmetric configurations for which $M^2$ is proportional to the
unique quadratic $U(4)$ invariant polynomial that can be constructed
from $\vec{Q}_e$ and $\vec{Q}_m$ \cite{osborn}. 

These results actually follow directly from the supersymmetry algebra, as we
now show. The ${\cal N}=4$ supersymmetry charges 
can be taken to be four two-component
complex spinors of $SL(2;\bC)$ in the fundamental ${\bf 4}$ representation of
$U(4)$. Let $Q_\alpha^i$ ($\alpha=1,2$, $i=1,2,3,4$) be these charges, with
$Q_{\dot\alpha\, i}$ their complex  conjugates in the $\bar {\bf 4}$
representation of $U(4)$. The matrix of anticommutators of these charges is
\be\label{42}
\{Q,Q\}= \pmatrix{\varepsilon_{\alpha\beta}Z^{ij}& 
\delta^i_k P_{\alpha\dot\alpha}\cr
\delta^l_j P_{\beta\dot\beta} & \varepsilon_{\dot\alpha\dot\beta}\bar Z_{kl}}
\ee
where $P$ is the 4-momentum and $Z$ a complex central charge in the ${\bf 6}$
representation of $SU(4)$. From the fact that the left-hand side of (\ref{42})
is positive semi-definite we deduce (by considering its determinant) the bound
\be\label{susyb}
M^4 - 2aM^2 + a^2- 4b \ge 0
\ee
where $M^2= -P^2$, and
\be
a= \frac{1}{4}\,Z^{ij}\bar Z_{ij} \,,\qquad a^2-4b= |{\rm Pf}\, Z^{ij}|^2
\ee
are $U(4)$-invariant polynomials (${\rm Pf}$ denotes the 
Pfaffian of the antisymmetric matrix $Z$; because this is complex 
one must take its modulus squared to get a polynomial $U(4)$
invariant). 
The bound (\ref{susyb}) is saturated when
\be
M^2 = a + 2\sqrt{b}
\ee
where we take the positive square root because this yields the strongest bound.
Comparison with (\ref{bpse}) shows that
\bea
a &=& |\vec{Q}_e|^2 + |\vec{Q}_m|^2\,, \nn
b &=& |\vec{Q}_e|^2|\vec{Q}_m|^2 - (\vec{Q}_e\cdot\vec{Q}_m)^2\,.
\eea
When $b=0$ this reduces to the formula $M^2= |\vec{Q}_e|^2 +
|\vec{Q}_m|^2$ applicable to 1/2 supersymmetric dyons. Otherwise only 1/4
supersymmetry is preserved.

\subsection{The flat background case}
We shall begin our discussion of the 1/4
supersymmetric dyons, and of their interpretation as string junctions,
by considering first the simpler case of a flat supergravity 
background, {\it i.e.} we set $H=1$.
Since both ${\bf E}$ and ${\bf B}$ are divergence-free, it follows from 
(\ref{equality}) that both $X$ and $Y$ are harmonic functions, vanishing
at worldspace infinity. As we will see, the most general string
junction can be realised by choosing $X$ and $Y$ to have two
centres. Hence, we choose two points $\bfsig_1$ and $\bfsig_2$ in
$\bE^3$ and solve the Laplace equations for $X$ and $Y$ everywhere else in
$\bE^3$ by setting
\bea
X &=& {q_1^X\over 4\pi |\bfsig-\bfsig_1|} + {q_2^X\over 4\pi
|\bfsig-\bfsig_2|}\nn 
Y &=& {q_1^Y \over 4\pi |\bfsig-\bfsig_1|}+ {q_2^Y \over 4\pi
|\bfsig-\bfsig_2|} 
\label{junction-flat}
\eea
where $q_1^X,q_2^X,q_1^Y$ and $q_2^Y$ are constants. 
The three strings in the string junction now arise from the
behaviour of the solution \eqn{junction-flat} in each of the three
regions of worldspace where it simplifies, namely near one of
the singularities or far away from both of them.  

Near $\bfsig=\bfsig_1$ we have
\be
X \sim {q_1^X \over 4\pi |\bfsig-\bfsig_1|}+X_0\,,\qquad
Y \sim {q_1^Y \over 4\pi |\bfsig-\bfsig_1|}+Y_0~.
\ee
where $X_0$ and $Y_0$ are constants.
In new coordinates rotated by an angle $\b$
\be
\pmatrix{X'\cr Y'} = \pmatrix{\cos\beta & \sin\beta \cr
-\sin\beta &\cos\beta} \pmatrix{X\cr Y}~,
\ee
with 
\be\label{angbeta}
\tan\b=q_1^Y/q_1^X~, 
\ee
we have 
\be
X' \sim {q\over 4\pi |\bfsig-\bfsig_1|}+X_0'\,,\qquad Y' \sim Y_0'~,
\ee
where 
\be\label{que}
q = \sqrt{{(q_1^X)}^2 + {(q_1^Y)}^2}\,.
\ee
The electric and magnetic fields are
\be
E \sim \cos(\alpha+\b) {\bf \nabla} X'\, , \qquad 
B \sim \sin(\alpha+\b) {\bf \nabla} X'\,~.
\ee
This looks like a 1/2 supersymmetric BIon with a spike along the 
$X'$-axis (see Fig. 1), total charge $q$ and
$\vartheta=\alpha+\b$. 
\begin{figure}
\label{junction}
\epsfbox{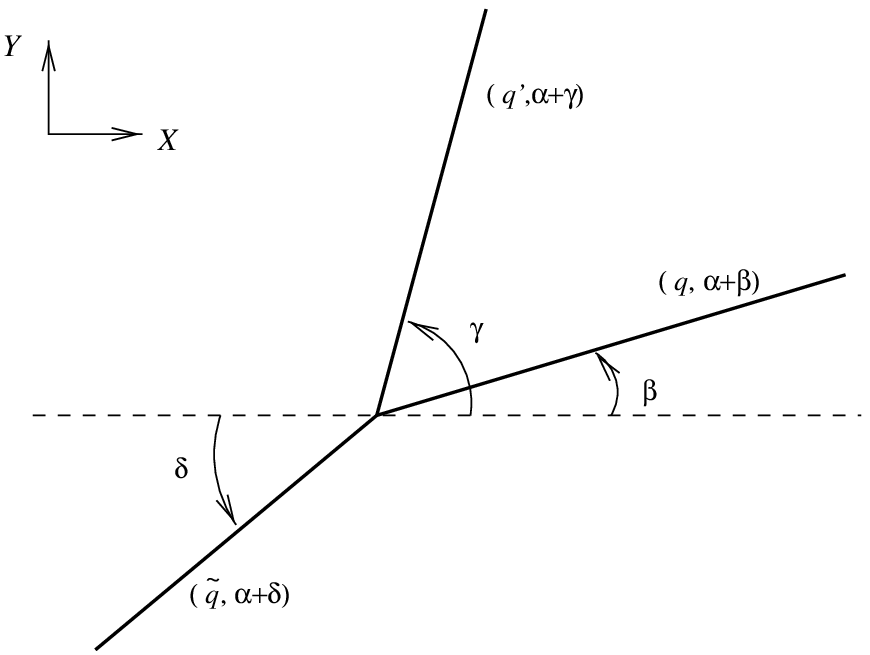}
\caption{The string junction described in the text.}
\end{figure}
In other words, it corresponds to an $(m,n)$ string
with 
\be\label{mandn}
m=q \,\cos(\a+\b) \, , \qquad n=q \, \sin(\a+\b)\,.
\ee 
Similarly, near $\bfsig =\bfsig_2$ we have
\be
X \sim {q_2^X\over 4\pi |\bfsig-\bfsig_2|} +X_0''\,,\qquad
Y \sim {q_2^Y\over 4\pi |\bfsig-\bfsig_2|}+ Y_0''~. 
\ee 
for constants $X_0''$ and $Y_0''$.
This again looks like a 1/2 supersymmetric BIon, now with
its spike at an angle $\gamma$ to the $X$-axis,
with total charge $q'$ and $\vartheta= \alpha+\gamma$ 
(see Fig. 1), where
\be\label{queprime}
q' = \sqrt{{(q_2^X)}^2 + {(q_2^Y)}^2}\,,\qquad 
\tan \gamma = q_2^Y/q_2^X~.
\ee
In other words, the two approximate BIon spikes
near either singularity are rotated relative to each other by the same
angle $\gamma- \beta$ in both space and charge space. 

Finally, consider a region far from both singularities, where
\be\label{joint}
X \sim {q_1^X+q_2^X\over 4\pi r}\, , \qquad
Y\sim {q_1^X + q_2^Y\over 4\pi r}\, ,
\ee
where $r$ is the distance from either singularity. 
We again have what looks like a 1/2 supersymmetric BIon, with its spike at
an angle $\delta$ to the $X$-axis, with total charge $\tilde{q}$ and
$\vartheta=\alpha+\delta$ (see fig. 1), where now
\be\label{faraway}
\tilde{q} = \sqrt{{(q_1^X + q_2^X)}^2 + {(q_1^Y+q_2^Y)}^2}
\ee
and
\be
\tan \delta = \frac{q_1^Y + q_2^Y}{q_1^X+q_2^X}~.
\label{far.away}
\ee 
Note that $\delta$ is defined by this formula only modulo $\pi$. In
this case we take the angle defining the orientation in charge space
to be $\alpha +\delta$ even though the angle defining the direction in
the $X-Y$ plane is $\alpha +\delta +\pi$. This is because we consider
the string orientations to be such that this string `enters' the 
junction whereas the other two `leave' it.

One important fact that supports the string junction interpretation of the 
configuration we have described is that, as is easily checked, it satisfies
both charge conservation (required for existence of a string junction
\cite{schwarz}) and tension balance (required for it to be static and
supersymmetric \cite{mukhi}).

\subsection{The two-centre background}
The energy of the configuration described in the previous
section is infinite, as
expected for a flat background.
To find finite energy solutions we return to the D3-brane background
with $H$ given by \eqn{Hwith2centres}. We set 
\bea
\vec{X}_1&=&(L_1,L_1',0,\ldots,0) \nn 
\vec{X}_2&=&(L_2,L_2',0,\ldots,0)~,
\eea
and we proceed according to the general
prescription given earlier. We remove two 3-balls from $\bE^3$ to create two
boundaries, $S_1$ and $S_2$, that are shells with the topology of 2-spheres. 
We take these to be
equipotentials with $(X,Y)=(L_1,L_1')$ on $S_1$ and 
$(X,Y)=(L_2,L_2')$ 
on $S_2$. The potential at infinity vanishes.
There is a unique solution to the Laplace equations for
$X$ and $Y$ subject to these boundary conditions.
Note that now the charges 
\bea
q_i^X &=& \oint_{S_i} d{\bf S}\cdot {\bf \nabla} X \,,\nn
q_i^Y &=& \oint_{S_i} d{\bf S}\cdot {\bf \nabla} Y\,,\qquad (i=1,2)
\eea
are fixed once the positions of the internal asymptotic regions
of the background are specified. This was to be expected for
the following reason. 
Because two of the strings are forced to go down each of these
regions, the background determines completely the relative orientation
among the three strings in the junction. The conditions of charge
conservation and tension balance then fix the values of the charges. 

The three strings arise again from the behaviour of the solution
near each of the shells and far away from both of them. Although the
final result is the same as in the flat case, 
the analysis is not so straightforward owing to the fact that the explicit 
expressions for $X$ and $Y$ are no longer available. However, we will
see that we have sufficient information to verify that the charge conservation
and tension balance conditions are still satisfied.

The main subtlety comes about when trying to assign a direction
to each of the strings. For definiteness, let us look at the region
near $S_1$. We have $X \sim L_1$ and $Y \sim L_1'$, so this region
is mapped into one of the internal asymptotic regions of the
background. We would like to determine the direction in the
$X-Y$ plane along which $X$ and $Y$ approach their asymptotic values
as we approach the surface $S_1$; this will be the direction along
which we will take the string to point. We can proceed in the 
following way. We consider approaching $S_1$ along a normal
to the surface, parametrised by some coordinate $r$ (see fig. 2).
\begin{figure}
\label{S1}
\epsfbox{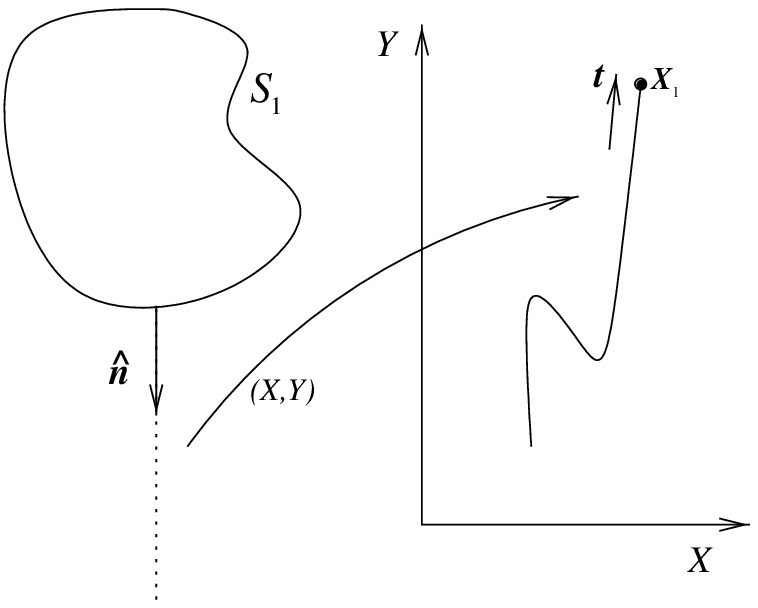}
\caption{The procedure described in the text to assign
directions to the strings.}
\end{figure}
\noindent
This normal is mapped to a curve in the $X-Y$ plane ending on
$\vec{X}_1$ (see Fig. 2). 
The tangent vector to this curve is $\vec{t}=(X',Y')$,
where the prime denotes differentiation with respect to $r$.
We still cannot take $\vec{t}$ to define the direction of the string,
because it depends on the choice of normal to the surface along which
we approach it. The most natural way to define the string direction
is therefore to take the average value $\vec{T}$ of $\vec{t}$ over
$S_1$. In other words, we define
\bea
T_1&\equiv&\oint_{S_1} dS\, X' 
= \oint_{S_1} d{\bf S}\cdot {\bf \nabla} X = q_1^X~, \nn
T_2&\equiv&\oint_{S_1} dS\, Y' 
= \oint_{S_1} d{\bf S}\cdot {\bf \nabla} Y = q_1^Y~,
\eea
where we made use of the fact that $\nax$ and $\nay$ are normal
to $S_1$ (in the electrostatic analogy, this corresponds to the 
electric field being normal to a perfectly conducting shell).
We take the string to lie at an angle $\b$ to the $X$-axis,
where 
\be
\tan \b =\frac{T_2}{T_1} = \frac{q_1^Y}{q_1^X}\, ,
\ee
which is the same as (\ref{angbeta}). 
The electric and magnetic charges carried by this string are
\bea
\oint_{S_1} d{\bf S}\cdot{\bf E} = \cos\a \,q_1^X -\sin\a\,q_1^Y
= q \, \cos(\a+\b)~, \nn
\oint_{S_1} d{\bf S}\cdot{\bf B} = \sin\a \,q_1^X +\cos\a\,q_1^Y
= q \, \sin(\a+\b)~, 
\eea
where $q$ is given in (\ref{que}). The string is therefore an $(m,n)$ string
with $m$ and $n$ as given in (\ref{mandn}). We conclude that the solution near
$S_1$ looks like  a 1/2 supersymmetric dyon of total charge $q$ and
$\vartheta=\a+\b$. 

Completely analogous arguments lead to the conclusion that the
solution near $S_2$ looks like another  1/2 supersymmetric dyon, this time with
its spike at an  angle $\gamma$ to the $X$-axis, of total charge $q'$ and
with $\vartheta=\a+\gamma$, where  $q'$ and $\gamma$ are given by
(\ref{queprime}). As in the flat background case, we see that the two
approximate 1/2 supersymmetric BIons are rotated by the same angle in space and
charge space. Finally, far away from both shells the solution behaves as
in \eqn{joint}, so it again looks like a third 1/2 supersymmetric
BIon with spike at an angle $\delta$ to the $X$-axis, 
total charge $\tilde{q}$ and $\vartheta=\a+\delta$, with 
$\tilde{q}$ and $\delta$ as in \eqn{faraway} and \eqn{far.away}.

In summary, the situation is again as depicted
in Fig. 1. Charge conservation and tension balance are 
satisfied. The difference is, of course, that
now the energy is finite.
 
\section{Finite energy M5-brane string solitons}

Infinite tension self-dual string solitons of the M5-brane worldvolume
field theory preserving 1/2 supersymmetry were found in \cite{HLW},
and shown to saturate a Bogomol'nyi-type bound in \cite{GGT}. Such
solutions reduce to BIons on a IIA D4-brane, and these are T-dual to the
electric D3-brane BIons discussed above. Here we shall find {\it
finite tension} self-dual strings on the M5-brane by considering an
M5-brane in the M5-brane supergravity background \eqn{m5sol}. We shall
follow the approach of \cite{GGT} for which the starting point is the
M5-brane Hamiltonian, which we now need in an M5-brane background.
This can be extracted from the Hamiltonian
form of the super M5-brane in a general superspace background given
in \cite{m5ham}, by proceeding along the lines of section 
\ref{d3ham}. In order to do this, one needs the expression
for the supergravity 6-form potential $C_{(6)}$ (dual to $C_{(3)}$),
which couples minimally to the M5-brane through the WZ term. For 
the M5-brane solution \eqn{m5sol} one choice is 
$C_{(6)}= U^{-1} \, vol(\bE^{(5,1)})$; other allowed choices differ 
by a gauge transformation and are therefore physically equivalent.
For this choice, and in the static gauge 
\be
X^{m} = (t,\sigma^a,\vec{X})\, ,
\ee
where $\sigma^a  (a=1,\ldots,5$) are the worldspace coordinates,
the energy density for static configurations is
\bea
({\cal E} + U^{-1})^2 &=& U^{-1/3} \big{[}
\det\,g + \frac{1}{2} \, g_{ab} \, g_{cd} \, \tilde{\cal{H}}^{ac}
\, \tilde{\cal{H}}^{bd} \nn
&& + U^{1/3} \, \delta^{ab} \, V_a V_b \big{]}~.
\eea
This is essentially the formula given in \cite{GGT} except for the
shift $U^{-1}$ in the energy due to the WZ term and the $U$-dependence
of the right hand side, which is part explicit and part implicit in
the $U$-dependence of the induced worldspace metric,
\be
g_{ab} = U^{-1/3} \, \delta_{ab} + U^{2/3}\, \pa_a\vec{X}\cdot\pa_b\vec{X}~.
\ee
Here, as in \cite{GGT},
\bea
\tilde{\cal{H}}^{ab} &=& \frac{1}{6} \, \varepsilon^{abcde}\, H_{cde} \nn
V_f &=& {1\over 24}\varepsilon^{abcde} H_{abc} \,H_{def}
\eea
where $H$ is a closed worldvolume 3-form. 

Our aim is now to find {\it finite tension} worldvolume self-dual string
solitons representing an M2-brane stretched between the parallel
test M5-brane and the background M5-brane. We therefore choose
a single-centre harmonic function $U$, with a singularity
on the $X$-axis at a distance $L$ from the origin in $\bE^5$.
Thus, on the $X$-axis we have
\be
U = 1 + \frac{Q}{(L-X)^3}~.
\ee
By arguments analogous to those of previous sections, the energy
will be minimised when all scalars but $X$ are zero. Setting these scalars
to zero, the expression for the energy then simplifies to
\bea
({\cal E} + U^{-1})^2 = U^{-2} &+& U^{-1}\, \left(
|\pa X|^2 + \frac{1}{2} \, |\tilde{\cal{H}}|^2 \right) \nn
&+& |\tilde{\cal{H}} \cdot \nax|^2 + |V|^2
\eea
where
\bea
|\tilde {\cal H}|^2 &=& \tilde {\cal H}^{ab} \tilde {\cal H}^{cd} \delta_{ac}
\delta_{bd}~, \nn 
|\tilde {\cal H}\cdot \partial X|^2 &=& 
\tilde {\cal H}^{ab}\tilde {\cal H}^{cd}\partial_b 
X\partial_d X\delta_{ac}~, \nn
|V|^2 &=& V_aV_b \delta^{ab}~,\nn
|\pa X|^2 &=& \pa_a X \pa_b X \delta^{ab}~.
\eea
This can be rewritten as
\bea
({\cal E}+U^{-1})^2 &=& \big|U^{-1}\, \zeta^a \pm 
\tilde {\cal H}^{ab}\partial_b X\big| ^2 \nn
&&+ 2 \, U^{-1} \, \big|\partial_{[a} X \zeta_{b]} 
\pm {1\over2} \delta_{ac}\delta_{bd} \tilde
{\cal H}^{cd}\big|^2 \nn
&& +\, U^{-1} \, (\zeta^a\partial_a X)^2 + |V|^2\,, 
\eea
where $\zeta$ is a unit length worldspace 5-vector, {\it i.e.}
\be
\zeta^a \zeta^b\delta_{ab} =1\, .
\ee
It follows that
\be
{\cal E} \ge |i_\zeta *_5 (dX\wedge H)|
\ee
where $*_5$ is the Hodge dual on worldspace, in the Euclidean metric,
and $i_\zeta$ denotes contraction with the (constant) 
vector field $\zeta$. This
inequality is saturated when
\be
H= \pm i_\zeta(*_5 dX) \, ,\qquad {\cal L}_\zeta X=0
\ee
where ${\cal L}_\zeta$ is the Lie derivative with respect to $\zeta$. These
conditions, which are the same as those of the flat background case, imply 
${\cal L}_\zeta H=0$. The minimum energy solution is therefore 
invariant under translations in the $\zeta$ direction, which we may take to be
compact with length $l$. The total energy is then
\be
{\it E} \ge l\times |Z|~,
\ee
where $Z$ is the topological charge found by integrating 
the 4-form $dX\wedge H$ over the 4-dimensional hypersurface $w_\zeta$ with 
normal $\zeta$, {\it i.e.}
\be
Z = \int_{w_\zeta} dX \wedge H = \int_{\pa w_\zeta} X \, H~ ,
\label{stringcharge}
\ee
where the last equality follows from the fact that $dH=0$. The absolute
value of $Z$ can be interpreted as the tension of the string soliton.

Because $H$ is closed, it also follows that $X$ is harmonic
on $w_\zeta$. For a single string with $X$ vanishing at transverse infinity
on the M5-brane we have
\be 
X=\frac{q}{2\pi^2 r^2} \,  \qquad  (r > r_0 \equiv \sqrt{q/2\pi^2 L})~,
\ee
where $r\equiv\sqrt{\sigma_1^2 + \ldots +\sigma_4^2}$ is the distance 
from the origin in $w_\zeta$. The lower bound on $r$ arises for reasons
analogous to those of section \ref{bions}: as $r$ approaches $r_0$ from above, 
$X$ approaches $L$ and 
\be
U \sim {Q\over [L-X(r)]^3} \sim {Qr_0^3 \over 8 L^3 u^3}\, .
\ee
where $u=r-r_0$. The asymptotic worldspace metric is
\be
ds^2 \sim C u(\zeta\cdot dx)^2  + Q^{2\over3} \left({du\over u}\right)^2
+ Cr_0^2 u\, d\Omega_3^2 
\ee
where $C=2L/Q^{1\over3} r_0$. The proper distance to $u=0$ is infinite so we
should restrict the worldspace coordinate $u$ to be positive. 
Note that not only does the 3-sphere at constant $u$ contract to a point at
$u=0$, but so also does the circle of length $l$ along the $\zeta$ direction.
Thus, it is not only the 3-sphere of radius $r_0$ that is mapped to the single
point $\vec{X} = (L,0,\ldots,0)$ in the target space; all points on the string
core at $r=r_0$ are also mapped to this point! 

We are now in a position to evaluate the integral of
\eqn{stringcharge} for the tension of the M5-brane string soliton. 
There are two components of $\partial
w_\zeta$. One is at $r=\infty$, where $X=0$, and the other is at
$r=r_0$, where $X=L$. The only contribution to the integral comes from
the latter boundary, so the string tension is
\be
Z=Lq
\ee
where 
\be
q \equiv \int_{S^3} \, H\, ,
\ee
is the string charge (as the string threads the 3-sphere). The total
energy is therefore $lLq$, as one would expect for a membrane of area
$lL$ and charge $q$. 

\section{Intersecting strings on the M5-brane}

We now turn to the study of two intersecting self-dual strings on 
the M5-brane, corresponding
to the spacetime configuration
$$
\begin{array}{lcccccccccc}
M5: & 1 & 2 & 3 & 4 & 5 & - & - & - & - & - \nn
M2: & - & - & - & 4 & - & 6 & - & - & - & - \nn
M2: & - & - & - & - & 5 & - & 7 & - & - & - 
\end{array}
$$ 
Therefore we allow for two scalars to be excited, in which case
\bea\label{enfive}
({\cal{E}}+U^{-1})^2 &=& U^{-2} + U^{-1}\big[|\pa X|^2 + |\pa Y|^2 +
\frac{1}{2}\,|\tilde{\cal{H}}|^2 \big]\nn
&+& {1\over2} |\pa X \wedge \pa Y|^2 
+ |\tilde{\cal{H}}\cdot\pa X|^2 + |\tilde{\cal{H}}\cdot\pa Y|^2 \nn
&+& (\pa X\cdot\tilde{\cal{H}}\cdot\pa Y)^2 + |V|^2
\eea
where
\bea
&& {1\over2} |\pa X \wedge \pa Y|^2 \equiv 
|\pa X|^2|\pa Y|^2 -( \pa X\cdot\pa Y)^2\, ,\nn
&& \pa X\cdot\tilde{\cal{H}}\cdot\pa Y \equiv
\pa_a X \, \tilde{\cal{H}}^{ab} \, \pa_b Y\, .
\eea

We are interested in minimum energy configurations associated with two
membranes that intersect the M5-brane in two non-parallel
directions, specified by two constant 5-vectors.
Let these vectors span the 4-5 plane and let $\sigma^\alpha$ 
($\alpha=1,2,3$) be the coordinates for the orthogonal complement of
worldspace. In this case, we have
\bea
V_\alpha &=& \epsilon_{\a\b\gamma}\, \tilde{\cal{H}}^{\b5} \,
\tilde{\cal{H}}^{\gamma4}
+\frac{1}{2}\,\epsilon_{\a\b\gamma}\, \tilde{\cal{H}}^{\b\gamma} \,
\tilde{\cal{H}}^{45} \nn
V_4&=& \frac{1}{2}\,\epsilon_{\a\b\gamma}\, \tilde{\cal{H}}^{\a\b} \,
\tilde{\cal{H}}^{5\gamma} \nn
V_5&=& -\frac{1}{2}\,\epsilon_{\a\b\gamma}\, \tilde{\cal{H}}^{\a\b} \,
\tilde{\cal{H}}^{4\gamma}~,
\eea
For notational convenience we define
\be
E^\a = \tilde{\cal H}^{\a5}\, ,\qquad
B^\a = \tilde{\cal H}^{\a4}
\ee
and
\be
K_\alpha = {1\over2}\varepsilon_{\a\b\c}\tilde{\cal H}^{\b\c}\, ,
\qquad \Pi = \tilde{\cal H}^{45}\, ,
\ee
so that
\bea
{\bf V} &=& {\bf E}\times {\bf B} + \Pi\, {\bf K}\nn
V_4&=& -{\bf E}\cdot{\bf K}\nn
V_5&=& {\bf B}\cdot{\bf K}
\eea
The right hand side of (\ref{enfive}) may now be rewritten as
\bea\label{mfdt}
&& U^{-2} + U^{-1}\big[ |\nax|^2 + |\nay|^2 + 
|{\bf E}|^2 + |{\bf B}|^2 \big]+ |\nax \times \nay|^2 \nn
&&+\, U^{-1}\big[\Pi^2 + (\pa_4 X)^2 + (\pa_5 X)^2 + 
(\pa_4 Y)^2 + (\pa_5 Y)^2 + |{\bf K}|^2]\nn
&& +\, |\pa_4 X \nay - \pa_4 Y \nax|^2
+ |\pa_5 X \nay - \pa_5 Y \nax|^2\nn
&& +\, (\pa_4 X\pa_5 Y - \pa_4 Y\pa_5 X)^2\nn
&& +\, ({\bf B}\cdot \nax - \Pi\pa_5 X)^2 +
({\bf E}\cdot \nax + \Pi\pa_4 X)^2 \nn
&& +\, ({\bf B}\cdot \nay - \Pi\pa_5 Y)^2 +
({\bf E}\cdot \nay + \Pi \pa_4 Y)^2 \nn
&& +\, |{\bf K} \times \nax + {\bf B}\pa_4 X + {\bf E}\pa_5 X|^2\nn
&& +\, |{\bf K} \times \nay + {\bf B}\pa_4 Y + {\bf E}\pa_5 Y|^2\nn
&& +\, |{\bf E}\times {\bf B} + \Pi{\bf K}|^2 + 
({\bf K}\cdot{\bf E})^2 + ({\bf K}\cdot{\bf B})^2 \nn
&& + \big[{\bf K}\cdot \nax \times \nay + \Pi(\pa_4 X \pa_5 Y -
\pa_4 Y \pa_5 X) \nn
&& +\, {\bf B}\cdot \nax\pa_4 Y - {\bf B}\cdot \nay \pa_4 X \nn
&& +\, {\bf E}\cdot \nax \pa_5 Y - {\bf E}\cdot \nay\pa_5 X \big]^2~.
\eea

We start to see the beginning of what looks like an expression for 
the energy of static configurations on the 3-brane obtained by 
compactifying the M5-brane on a 2-torus, as expected from the
standard duality between M-theory and the IIB superstring theory.
The comparison is, however, complicated by several factors.
One is that the notion of `static' for an M5-brane configuration does
not coincide with what we meant by this term in the D3-brane case. The
point is that the 3-form $H$ on the M5-brane worldspace includes
configuration space variables {\it and} their conjugate momenta (via
a constraint relating $\tilde {\cal H}$ to the momentum conjugate to
the 2-form potential \cite{m5ham}). A related point is that the space
transverse to the M5-brane is only 5-dimensional whereas the space
transverse to the D3-brane is 6-dimensional. The `sixth' scalar, and its
conjugate momentum are encoded in the 3-form $H$. In fact, one can
roughly view ${\bf K}$ as the field strength of the sixth scalar and 
$\Pi$ as its conjugate momentum, although ${\bf K}$ is only a closed
1-form when ${\bf E}$ and ${\bf B}$ are independent of $\sigma^4$ and
$\sigma^5$. Clearly, a direct comparison with the 
D3-brane energy would require that we include in the latter some
`3-scalar' terms as well as some terms that vanish for static
configurations (in the D3-brane sense). Instead, we shall set ${\bf K}$ and
$\Pi$ to zero and compare with the expression for the energy of
static `2-scalar' D3-brane configurations. Note that, under these
conditions, the closure of the 3-form $H$ implies that 
${\bf E}$ and ${\bf B}$ are divergence-free, in which case they may
be identified as the electric and magnetic fields on the D3-brane. 

With this simplification,
the expression (\ref{mfdt}) can be written, for arbitrary angle
$\varphi$, as
\bea
&&\big[U^{-1} + \cos\varphi\, {\bf E}\cdot \nax - 
\sin\varphi\, {\bf E}\cdot \nay \nn
&&+\, \sin\varphi\, {\bf B}\cdot \nax + \cos\varphi\, {\bf B}\cdot
\nay\big]^2 \nn
&&+ |{\bf E}\times {\bf B} - \nax\times\nay|^2 +
 (\pa_4 X\pa_5 Y - \pa_4 Y\pa_5 X)^2\nn
&& +\,  U^{-1}\big[ (\pa_4 X)^2 + (\pa_5 X)^2 + 
(\pa_4 Y)^2 + (\pa_5 Y)^2 \big]\nn
&& +\, |\pa_4 X \nay - \pa_4 Y \nax|^2
+ |\pa_5 X \nay - \pa_5 Y \nax|^2\nn
&& +\, (\pa_4 X\pa_5 Y - \pa_4 Y\pa_5 X)^2\nn
&& +\,U^{-1}|{\bf E} -\cos\varphi\,\nax +\sin\varphi\, \nay|^2\nn 
&&+\, U^{-1}|{\bf B} -\sin\varphi\,\nax -\cos\varphi\, \nay|^2 \nn
&& +\, \big[ \sin\varphi\, {\bf E}\cdot \nax + \cos\varphi\, {\bf E}\cdot
\nay \nn
&&-\, \cos\varphi\, {\bf B}\cdot \nax + 
\sin\varphi\, {\bf B}\cdot \nay\big]^2\nn
&& +\, |{\bf B}\,\pa_4 X + {\bf E}\,\pa_5 X|^2 + 
|{\bf B}\,\pa_4 Y + {\bf E}\,\pa_5 Y|^2\nn
&& +\, \big[{\bf B}\cdot \nax\pa_4 Y
- {\bf B}\cdot \nay \pa_4 X \nn
&&\qquad +\, {\bf E}\cdot \nax \pa_5 Y - 
{\bf E}\cdot \nay\pa_5 X \big]^2\, .
\eea
We thereby deduce the bound
\bea\label{mfivebound}
{\cal E} &\ge& \cos\varphi\, {\bf E}\cdot \nax - \sin\varphi {\bf
E}\cdot \nay\nn
&&\qquad  
+\,  \sin\varphi\, {\bf B}\cdot \nax + \cos\varphi\, {\bf B}\cdot \nay\, ,
\eea
for any $\varphi$, with equality when both
\be
\pa_4 X=\pa_5 X=\pa_4 Y=\pa_5 Y =0\, ,
\ee
and
\bea
{\bf E} &=& \cos\varphi\, \nax -\sin\varphi\, \nay\nn
{\bf B} &=& \sin\varphi\, \nax + \cos\varphi\, \nay\, .
\eea
These conditions are precisely those found 
in \cite{GLW} to be associated with 1/4 supersymmetry. When they
are satisfied, the 3-vectors ${\bf E}$ and ${\bf B}$ are independent of
$\sigma^4$ and $\sigma^5$. The energy of the minimal energy
configuration on the $T^2$-wrapped M5-brane is therefore proportional
to the expression (\ref{bpse}) for the energy of a 1/4 supersymmetric
dyon on the D3-brane. 

The 1/4 supersymmetric dyons discussed earlier thus aquire an M-theory
interpretation as intersections on the M5-brane of two string
boundaries of two M2-branes. When the two M2-branes intersect
orthogonally, so do the strings. This maps to a IIB configuration 
composed of two orthogonal BIon spikes that are also othogonal in
charge space. A rotation of the M2-branes away from orthogonality 
preserving 1/4 supersymmetry corresponds to a simultaneous rotation 
(of one membrane relative to the other) by the same angle in the 
M5-brane and in the space transverse to it.
This corresponds in the IIB theory to a simultaneous rotation by the
same angle in transverse space and charge space. 

We can now find finite energy 1/4 supersymmetric solitons on the
M5-brane by the procedure described earlier, but we should mention 
that the relation between the 1/4 supersymmetric 
M5-brane solitons and the string-junction dyons on the D3-brane is 
straightforward only in the case of the infinite energy solutions, in 
a flat background, because in
general the harmonic function $U$ of the M5-brane background is not
the same as the harmonic function $H$ of the D3-brane background. The
M5-brane transverse space is only five-dimensional whereas 
it is six-dimensional for the D3-brane. 

We should also mention that there is more than one possible 
interpretation of D3-brane dyons as M5-brane solitons. Recall that the
6-space transverse to the D3-brane is reduced to a 5-space transverse
to the M5-brane. This leads one to wonder what the M-theory interpretation is
of a D3-brane BIon with its spike in this sixth dimension. If the
D3-brane BIon is purely magnetic then the answer
is that it corresponds to a marginal bound state of the M5-brane with
an M-wave travelling along it. If we compactify along the direction of
the wave then we get a D0-brane on a D4-brane of IIA superstring
theory. This has a T-dual in IIB superstring theory as a D-string
ending on a D3-brane, which is a magnetic BIon from the perspective
of the D3-brane worldvolume. One gets an electric D3-brane BIon from
the same M-theory configuration by reducing along another direction in
the M5-brane orthogonal to the M-wave. This leads to a wave on a
D4-brane. T-dualizing along the wave direction then yields the desired
configuration.

\section{Discussion}

We have seen that finite energy configurations of IIB superstring theory in
which $(m,n)$ strings are suspended between D3-branes 
can be found as 1/2 or 1/4
supersymmetric solitons on a test D3-brane in a supergravity D3-brane
background, the 1/4 supersymmetric solitons having a natural interpretation in
terms of string junctions. These solitons are abelian analogues of the finite
energy supersymmetric  solitons of $D=4$ ${\cal N}=4$ 
SYM theory. There are similar 1/2 and 1/4
supersymmetric solitons on a test M5-brane in a supergravity M5-brane
background, related by dualities to those on the D3-brane. In our exposition of
these results we glossed over a few points, and we shall conclude with a brief
discussion of them. 

The first point has to do with whether the BIon solutions we have have found
are really non-singular. Consider the one BIon solution in which the sphere in
$\bE^3$ of radius $r_0$ is mapped by $X(r)$ to the `centre' $X=L$ of
the background metric. The induced worldspace metric near $r=r_0$
was given as $ds_3^2$ in (\ref{nearsing}), with $u=r-r_0$. The corresponding
worldvolume metric is, for $Q=1$ and after a constant rescaling of the time
coordinate,
\be\label{full}
ds^2_4 = - u^2 dt^2 + \left({du\over u}\right)^2 + L^2 u^2
d\Omega_2^2\, .
\ee
The submanifold with $d\Omega^2_2=0$ is just $adS_2$, with $u=0$ a Killing
horizon of $\partial_t$. This is at infinite proper distance on spacelike
geodesics of constant $t$ but at {\it finite} affine parameter on timelike
or null geodesics. This is not unexpected because the singularity of the
spacetime metric at its centre is also a Killing horizon at a finite affine
parameter on timelike or null geodesics. However, although the singularity at
$u=0$ is just a coordinate singularity of $adS_2$, it 
is a curvature singularity
of the full 4-metric (\ref{full}) because the 2-spheres at constant $u$ shrink
to points at $u=0$. 

It is not clear to us whether this is really a problem 
because there are no test particles on the D3-brane 
moving on timelike geodesics
in this worldvolume metric. If we were to consider the time dependent problem
in which a wave on the D3-brane scatters from a BIon then we would have to
first solve the DBI equations for this problem and then recompute the 
worldvolume metric. In general, one would expect the result to differ from
the static BIon metric of (\ref{full}), which would be of direct
relevance to the time-dependent problem only if all scalar
fluctuations were to vanish. It seems unlikely that the DBI
equations will have such solutions when linearized about
the static  BIon configuration (rather than about the Minkowski vacuum) but we
have not investigated this in detail. The whole area of time-dependent
scattering solutions involving BIons deserves a separate study. 

A second point concerns the string junction interpretation of the 1/4
supersymmetric DBI dyons on the D3-brane. Recall that this involves, in the
case of a flat background, singularities of two harmonic functions $X$ and $Y$
at points with $\bE^3$ coordinates $\bfsig_1$ and $\bfsig_2$. Near either
singularity, and far from both, the solution is approximately that of a single
1/2 supersymmetric BIon. The term `far from' here refers to a region in
which the distance from either singularity (in the $\bE^3$ metric) is much
larger than $|\bfsig_1-\bfsig_2|$. The larger $|\bfsig_1-\bfsig_2|$ is, the
further out is this region and the smaller is the deviaton of the worldspace
metric from the Euclidean metric. Thus, for large $|\bfsig_1-\bfsig_2|$ it is
more natural to interpret the worldspace configuration as one for which two
strings meet the D3-brane at widely separated points. Only as the separation of
these points decreases does the string junction interpretation become the
natural one. Even in this case one could interpret the deviation of the
worldspace from the Euclidean metric as that required to support two strings
meeting the D3-brane at the points $\bfsig=\bfsig_1$ and 
$\bfsig=\bfsig_2$, just
as the single BIon solution has the alternative interpretation as the deviation
required to support an attached string \cite{Gib,hash}, 
rather than as the string itself.

Finally, we should mention that string junction dyons constitute a
special case of string web dyons \cite{kol} in which three or more parallel
D3-branes are connected by a network of strings meeting at string junctions
\cite{sen}. The logic of our approach would suggest that these should also
have a worldvolume interpretation but we have not been able to verify this.
In the first instance one could seek new infinite energy solutions in which
strings arriving from other branes at infinity are realized as point charges.
The topological features of the network might then be encoded in branch cuts,
but it is unclear to us whether this makes sense, and if so how it works in
detail.

\vskip 0.5cm


{\bf Acknowledgments:}
D.M. thanks colleagues at DAMTP for the hospitality during
the time this work was performed. D.M. is supported by a fellowship from
the Comissionat per a Universitats i Recerca de la Generalitat de
Catalunya. J.P.G. thanks EPSRC for financial support.  We thank Fay
Dowker, Gary Gibbons and George Papadopoulos for helpful 
conversations, and Barak Kol for raising the issue of string webs. 



\bigskip

\begin{thebibliography}{99}

\bibitem{hhs}
K. Hashimoto, H. Hata and N. Sasakura, {\sl 3-string junctions and BPS
saturated solutions in $SU(3)$ supersymmetric Yang-Mills theory},
Phys. Lett. {\bf 431B} (1998) 303;\\
{\sl Multi-pronged strings and BPS saturated states in $SU(N)$
supersymmetric Yang-Mills theory}, Nucl. Phys. {\bf B535} (1998) 83.

\bibitem{ko}
T. Kawano and K. Okuyama, {\sl String networks and 1/4 BPS states in
N=4 $SU(n)$ supersymmetric Yang-Mills theory}, Phys. Lett. {\bf 432B}
(1998) 338.

\bibitem{lee}
K. Lee and P. Yi, {\sl Dyons in N=4 supersymmetric theories and three-pronged
string}, Phys. Rev. {\bf D58} (1998) 066005.

\bibitem{berg}
O. Bergman, {\sl Three-pronged strings and 1/4 BPS states in N=4 super
Yang-Mills theory}, Nucl. Phys. {\bf B525} (1998) 104.

\bibitem{schwarz}
J.H. Schwarz, {\sl Lectures on superstring and M-theory dualities},
Nucl. Phys. Proc. Suppl. {\bf B55} (1997) 1.

\bibitem{mukhi}
S. Mukhi and S. Dasgupta, {\sl BPS nature of 3-string junctions},
Phys. Lett. {\bf 423B} (1998) 261.

\bibitem{GGT}
J.P. Gauntlett, J. Gomis and P.K. Townsend, {\sl BPS bounds for worldvolume
branes}, JHEP 01 (1998) 003.

\bibitem{brecher}
M.J. Perry and D. Brecher, {\sl Bound states of D-branes and the
non-Abelian Born-Infeld action}, Nucl. Phys. {\bf B527} (1998) 121;\\
D. Brecher, {\sl BPS states of the non-abelian Born-Infeld action}, Phys. Lett.
{\bf 442B} (1998) 117.

\bibitem{CM}
C. Callan and J. Maldacena, {\sl Brane dynamics from the Born-Infeld action},
Nucl. Phys. {\bf B513} (1998) 198.

\bibitem{Gib}
G.W. Gibbons, {\sl Born-Infeld particles and Dirichlet p-branes},
Nucl. Phys. {\bf B514} (1998) 603.

\bibitem{GP}
G. Papadopoulos and J. Gutowski, {\sl The dynamics of D3-brane dyons and toric
hyper-Kahler geometry}, hep-th/9811207.

\bibitem{HLW}
P.S. Howe, N.D. Lambert and P.C. West, {\sl The self-dual string
soliton}, Nucl. Phys. {\bf B515} (1998) 203.

\bibitem{GLW}
J.P. Gauntlett, N.D. Lambert and P.C. West, {\sl Supersymmetric five-brane
solitons}, hep-th/9811024. 

\bibitem{BT}
E. Bergshoeff and P.K. Townsend, {\sl Super D-branes revisited},
Nucl. Phys. {\bf B531} (1998) 226. 


\bibitem{hollowood}
C. Fraser and T.J. Hollowood, {\sl Semi-classical quantization in N=4
supersymmetric Yang-Mills theory and duality}, Phys. Lett. {\bf 402B}
(1997) 106.

\bibitem{osborn}
H. Osborn, {\sl Topological charges for N=4 supersymmetric gauge
theories and monopoles of spin 1}, Phys. Lett. {\bf 83B} (1979) 321. 

\bibitem{m5ham}
E. Bergshoeff, D. Sorokin and P.K. Townsend,
{\sl The M5-Brane Hamiltonian}, Nucl. Phys. {\bf B533} (1998) 303.

\bibitem{hash}
A. Hashimoto, {\sl The shape of branes pulled by strings}, Phys. Rev. {\bf D57}
(1998) 6441.

\bibitem{kol}
O. Bergman and B. Kol, {\sl String webs and 1/4 BPS monopoles}, 
Nucl. Phys. {\bf B536} (1998) 149.

\bibitem{sen}
A. Sen, {\sl String networks}, JHEP 9803: 005 (1998)

  
 

\end{thebibliography}
\end{document}